\documentstyle[aps,prl,epsfig,multicol]{revtex}
\input epsf

\begin{document}
\title{Mechanisms of Electrical Conductivity in $Y_{1-x}Ca_{x}Ba_{2}Cu_{3}O_{6.1}$ System}
\author{ P. Starowicz, and A. Szytu\l{}a }
\address{M. Smoluchowski Institute of Physics, Jagiellonian University, Reymonta 4, 30-059
Krak\'{o}w, Poland\\
\vspace{0.2cm} 
}
\maketitle
\begin{abstract}
Systematic studies of transport properties in deoxygenated
$Y_{1-x}Ca_{x}Ba_{2}Cu_{3}O_{6.1}$ series allowed to propose a
diagram of conductivity mechanisms for this system. At
intermediate temperature a variable range hopping (VRH) in 2
dimensions prevails. At lower temperature VRH in the presence of a
Coulomb gap for smaller x and VRH in 2 dimensions for larger x are
found. In a vicinity of superconductivity we observe conductivity
proportional to $\sqrt{T}$. Thermally activated conductivity
dominates at higher temperature. This diagram may be universal for
the whole family of undoped high $T_{c}$ related cuprates.
\end{abstract}
\pacs{71.30.+h,72.20.Ee,74.62.Dh,74.72.Bk}

\begin{multicols}{2}
\narrowtext High $T_{c}$ superconductors and their parent
compounds remain ones of the most intriguing and not completely
understood systems despite an impressive quantity of collected
experimental material. All the different families of
superconducting cuprates exhibit similar behavior as a function of
holes doping to Cu-O planes\cite{Presland,d8,d3}. Like in general
case of the high $T_{c}$ related cuprates the undoped,
deoxygenated $YBa_{2}Cu_{3}O_{6.1}$ is an antiferromagnetic
insulator and increase in oxygen content provides holes to Cu-O
planes, destroys static antiferromagnetic order, forces the
insulator to metal transition and induces
superconductivity\cite{d2}. There is a considerable number of
reports on the effect of oxygen doping to the
$YBa_{2}Cu_{3}O_{6+d}$ system\cite{d3,d2}. An alternative method
is realized by divalent Ca substitution for trivalent Y, while the
oxygen content remains constant\cite{d8,d4,d6,d7}. This chemical
substitution resulting in the injection of holes into Cu-O planes
allows to manipulate in carrier concentration quite precisely
within the limit of Ca solubility. The deoxygenated
$Y_{1-x}Ca_{x}Ba_{2}Cu_{3}O_{6.1}$ system exhibits
superconductivity for higher Ca level\cite{d8,d4,d6,d7} and is
characterized by a phase diagram\cite{d7} analogous to the diagram
for $YBa_{2}Cu_{3}O_{6+d}$ with variable oxygen content\cite{d2}.

A number of studies concerning electrical conductivity have been
performed in order to understand the transport properties of the
high $T_{c}$ related compounds and to investigate the phenomena
occurring in the insulator to metal
transition\cite{d44,d46,d56,d45,d55,d57,d43,d54}. For highly
doped, superconducting samples a linear $\rho(T)$ (resistivity
versus temperature) dependence with a positive slope in a normal
state has been found\cite{d3}, whereas for less doped samples
different kinds of variable range hopping (VRH) conductivity have
been observed\cite{d44,d46,d56,d45,d55,d57}. Electrical
resistivity in the VRH mechanism follows the expression:
\begin{eqnarray}{\rho}=\rho_{0}\exp(T/T_{0})^{-n}\label{E1}\end{eqnarray}
where $n=1/4$ stands for VRH in 3 dimensions\cite{d31} observed
e.g. in $Bi_{2}Sr_{2}Ca_{1-x}Y_{x}Cu_{2}O_{8+y}$
ceramics\cite{d44}, $Bi_{2}Sr_{2}Ca_{1-x}Gd_{x}Cu_{2}O_{8+d}$
ceramics\cite{d46}, both single crystals and ceramics of
$La_{2-y}Sr_{y}Cu_{1-x}Li_{x}O_{4-\delta}$\cite{d56},
$La_{2-x}Sr_{x}CuO_{4}$ ceramics\cite{d45}, $La_{2}CuO_{4+\delta}$
single crystals\cite{d55} or $Y_{1-x}Pr_{x}Ba_{2}Cu_{3}O_{7}$
crystals\cite{d57}. The exponent $n$ equals $1/2$ for the VRH
conductivity in the presence of a Coulomb gap or carrier
interactions\cite{d48} which was found e. g. in
$La_{2-x}Sr_{x}CuO_{4}$\cite{d45}. The exponent $n=1/3$ is a
prediction for VRH in 2 dimensions\cite{d31}, observed e.g. in
$Bi_{2}Sr_{2}Ca_{1-x}Y_{x}Cu_{2}O_{8+y}$ ceramics\cite{d44} or
$Bi_{2}Sr_{2}Ca_{1-x}Gd_{x}Cu_{2}O_{8+d}$ ceramics\cite{d46}. It
was also found that thermally activated conductivity ($n=1$) is
responsible for transport properties at higher temperature in
$Bi_{2}Sr_{2}Ca_{1-x}Y_{x}Cu_{2}O_{8+y}$ ceramics\cite{d44} or in
$La_{2}CuO_{4+\delta}$ single crystals\cite{d55}. The theoretical
predictions for disordered metals\cite{d50} propose the expression
\begin{eqnarray}\sigma=\sigma_{0}+m\cdot\sqrt{T}\label{E2}\end{eqnarray}
where $\sigma$ is static electrical conductivity. This relation
was observed in $La_{1.85}Sr_{0.15}CuO_{4}$ ceramics with
impurities\cite{d43}. Conductivity following the relation
\begin{eqnarray}\sigma\propto\ln(T)\label{E3}\end{eqnarray}
indicates a weak localization in 2 dimensions or electron -
electron interactions in the presence of disorder in 2
dimensions\cite{d50}, found in $Nd_{1.85}Ce_{0.15}CuO_{4-\delta}$
in ab plane\cite{d54}. Luttinger liquid theory in two
dimensions\cite{d49} predicts the following behavior of electrical
resistivity:
\begin{eqnarray}{\rho}\propto T^{1-4\alpha},\label{E4}\end{eqnarray}
where $0\leq\alpha\leq\frac{1}{2}$ and $\alpha$ decreases in the
transition from insulator to metal. To get a better insight into
the conductivity mechanisms and systematize our knowledge
concerning the insulator to metal and superconductor transition we
performed measurements of electrical resistivity on a series of
$Y_{1-x}Ca_{x}Ba_{2}Cu_{3}O_{6.1}$ samples.

$Y_{1-x}Ca_{x}Ba_{2}Cu_{3}O_{6+d}$ samples were prepared by the
standard ceramic method. Subsequently, they were subjected to
reduction in flowing Argon at 730 $^{o}C$ for 24 hours followed by
slow cooling to room temperature. The oxygen content in the
obtained samples amounted between 6.08 and 6.16. More details
concerning preparation and characterization of the samples are
given elsewhere\cite{Starowicz}. Electrical resistivity
measurements were performed with the four point contact method
with reversing current direction in the temperature range of 6 K -
300 K. Electrical contacts were made of InAu alloy on mechanically
cleaned sample surface. At each temperature, measurements with
different electrical currents were made to verify if the
resistivity is ohmic and to exclude the effect of heating or other
non-linear phenomena. The applied current varied from 3 mA to the
values lower than 1 $\mu A$ for the samples with the highest
resistivity.

The measurements of electrical resistivity revealed a gradual
modification of the transport properties with doping. As presented
in Fig. 1 the system is insulating for x=0.00 and Ca substitution
successively diminishes electrical resistivity. This is well
reflected in the resistivity dependence as a function of Ca
content at constant temperature (Fig. 1, inset). Superconductivity
appears at x=0.3 with the onset temperature equal to 35 K, whereas
is not visible above 6 K in the resistivity curve for x=0.26. The
quantity of calcium at which superconductivity is induced was
different than in the other reports\cite{d8,d6,d7}. The
discrepancy in this field may be due to different methods of
sample preparation, which can influence the existence of disorder
or alter the ratio of Ca substituted into Y to substituted in Ba
lattice positions. For example, the quoted reports\cite{d8,d6}
concern the samples reduced by quenching in contrary to our method
of preparation.

We will not discuss the metallic side of a metal-insulator
transition, for which a linear $\rho(T)$ dependence is typically
observed in high $T_{c}$ related cuprates. However, an ambiguous
situation in the insulating side gave us motivation to verify the
listed models of conductivity by drawing the data in appropriate
scales. The plots of $\log(\rho)$ versus $T^{-\frac{1}{2}}$,
$T^{-\frac{1}{3}}$ and $T^{-\frac{1}{4}}$ are presented in the
Fig. 2 to consider if the VRH mechanisms occur. With the
superficial analysis of the Fig. 2 it is difficult to confirm or
exclude the existence of the mentioned mechanisms. In order to
resolve which particular mechanism can dominate in different
temperature and doping regions a normalized $\chi^{2}$ test was
applied as the discussed relations were fitted to the experimental
data. Different forms \cite{refT} of the expression (\ref{E1}):
$\rho=\rho_{0}\exp((T/\tau)^{-1})$,
$\rho=\rho_{0}'\exp((T/T_{0}')^{-\frac{1}{2}})$,
$\rho=\rho_{0}''\exp((T/T_{0}'')^{-\frac{1}{3}})$,
$\rho=\rho_{0}'''\exp((T/T_{0}''')^{-\frac{1}{4}})$ and the power
function (\ref{E4}) were fitted in different temperature ranges,
particularly those, with visible linearity in appropriate scale.
The relations (\ref{E2}) and (\ref{E3}) were fitted for
$x\geq0.16$, as for lower x they do not match the data visually.
There was more than ten fitting ranges of temperature for each
sample. In the Table 1, the fits with the lowest normalized
$\chi^{2}$ values are indicated for the given temperature ranges.
The proposed ranges in the Table 1 represent regions, where
selected models of resistivity seem to be prevalent. This is
reflected in larger differences between the smallest $\chi^{2}$
values for the proposed model and $\chi^{2}$ for the others.
\begin{figure}[c!]
    \vspace{0.5cm}
    \centerline{\epsfig{file=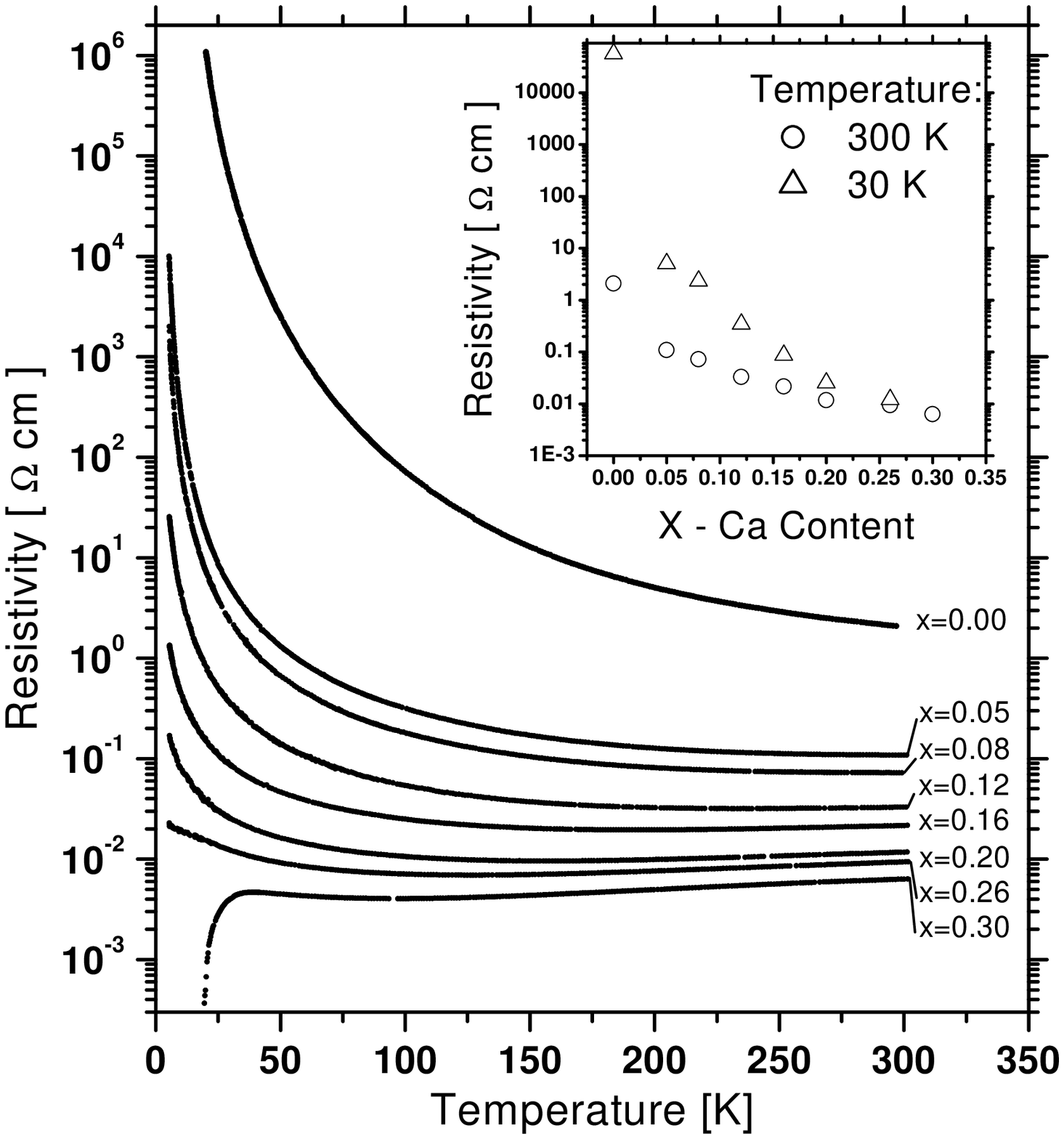,bbllx=15,bblly=230,bburx=534,bbury=780,width=7.5cm}}
    \vspace{0.3cm}
    \caption[fig1]{Electrical resistivity versus temperature for $Y_{1-x}Ca_{x}Ba_{2}Cu_{3}O_{6.1}$ samples.
Inset: The resistivity at constant temperature as a function of Ca
concentration}
\end{figure}
Before starting a discussion of the results from the Table 1 we
will examine conditions for the existence of VRH\cite{d31,d48}. At
the considered range of temperature the most probable hopping
distance $R$ should be higher than the localization length $\xi$
\begin{eqnarray}R>\xi,\label{E5}\end{eqnarray}
The evaluation of the inequality (\ref{E5}) leads to $T_{0}'/4>T$,
$T_{0}''/27>T$ and $T_{0}'''/50>T$ respectively for VRH in the
presence of a Coulomb gap, VRH in two dimensions, and VRH in three
dimensions. Hence, the parameters $T_{0}'$, $T_{0}''$ and
$T_{0}'''$ set necessary conditions for the temperature at which
VRH can be accepted.

The analysis of the data from the Table 1 allowed to find some
regularities in a behavior of electrical resistivity and we
propose a diagram of the conductivity mechanisms for this system
in the Fig. 3. It is postulated that different mechanisms are
dominating in some temperature and doping regions. In reality, the
different processes must coexist in certain areas of the diagram.
Thus, it is purposeless to draw borders between the regions.

For the majority of investigated systems, we can distinguish three
temperature ranges. In the intermediate range the existence of VRH
in 2 dimensions ($n=1/3$) can be proposed. These regions are 26 K
- 80 K for the specimen without Ca, 12 K - 70 K for samples with
$x=0.05$ to 0.20 and 20 K to 80 K for the sample with $x=0.26$.
The assumption (\ref{E5}) is fulfilled for the systems with
$x=0.00$ to $x=0.16$ at the considered temperature. For the sample
with $x=0.20$ VRH in 2 dimensions could be the case only below 25
K but at this temperature VRH in three dimensions will be
prevalent ( Table 1 ). The condition (\ref{E5}) is failed for the
systems with $x>0.20$.
\begin{figure}[c!]
    \centerline{\epsfig{file=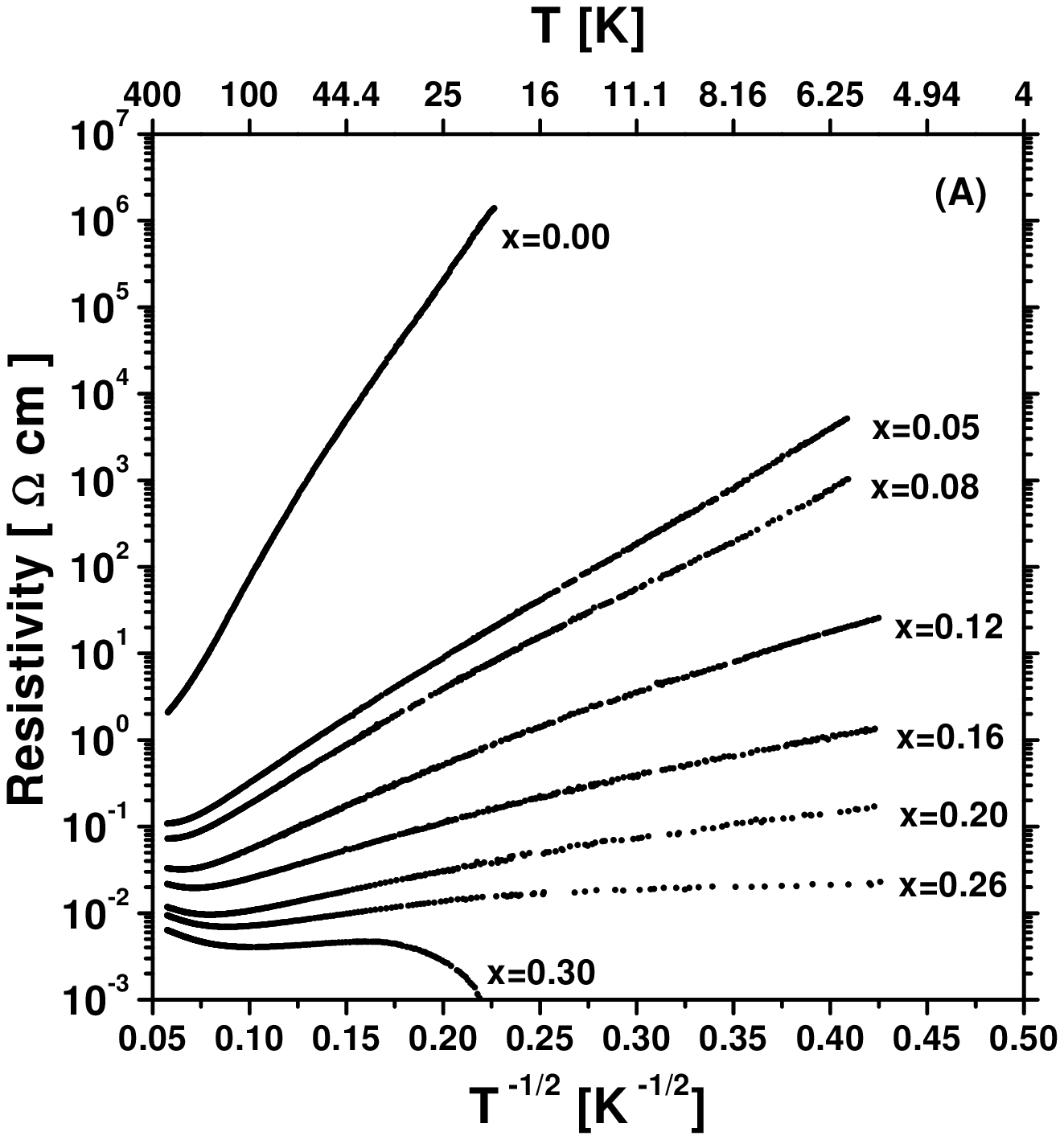,bbllx=60,bblly=380,bburx=476,bbury=800,width=6cm}}
    \vspace{0.8cm}
    \centerline{\epsfig{file=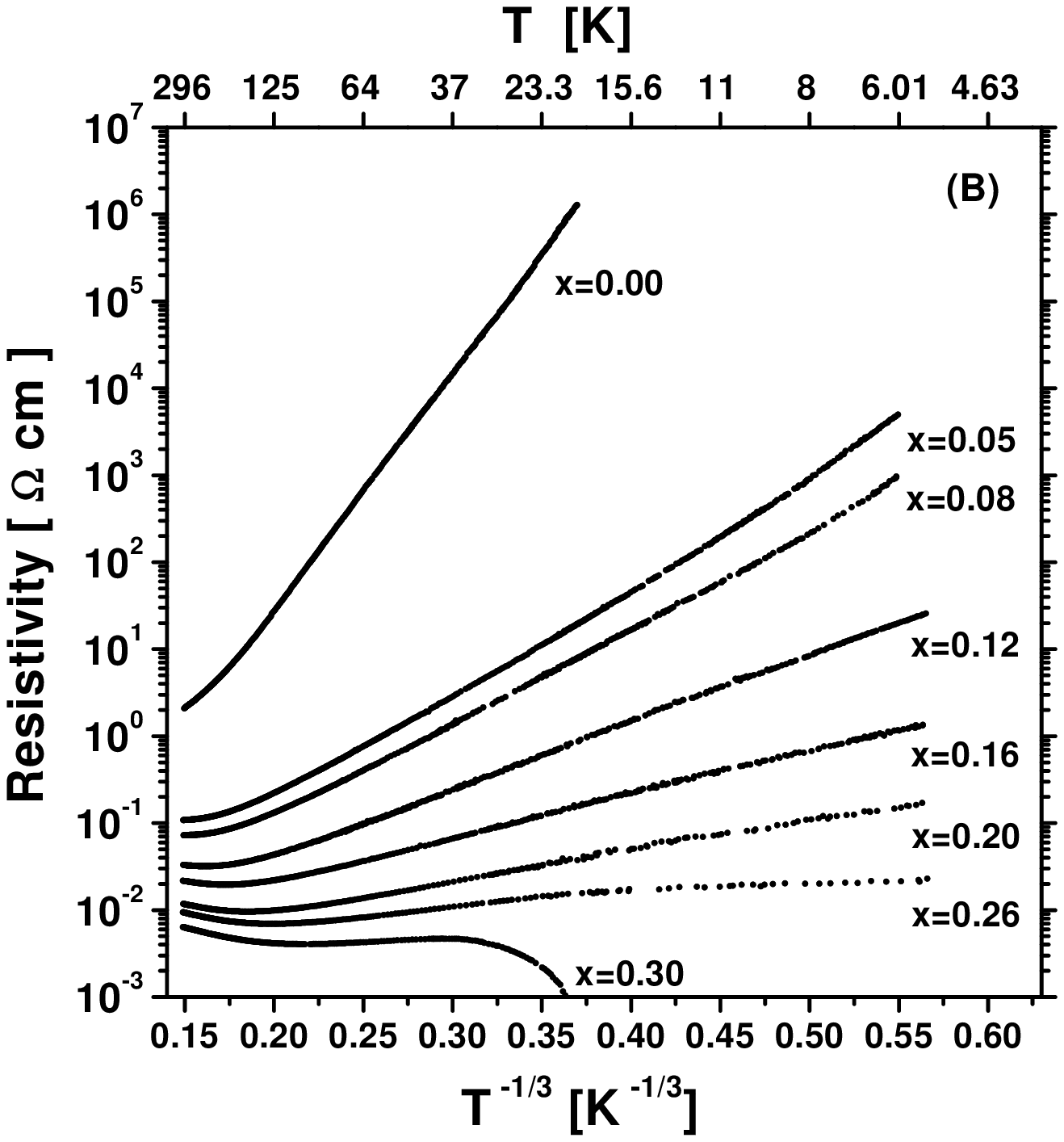,bbllx=60,bblly=380,bburx=476,bbury=800,width=6cm}}
    \vspace{0.8cm}
    \centerline{\epsfig{file=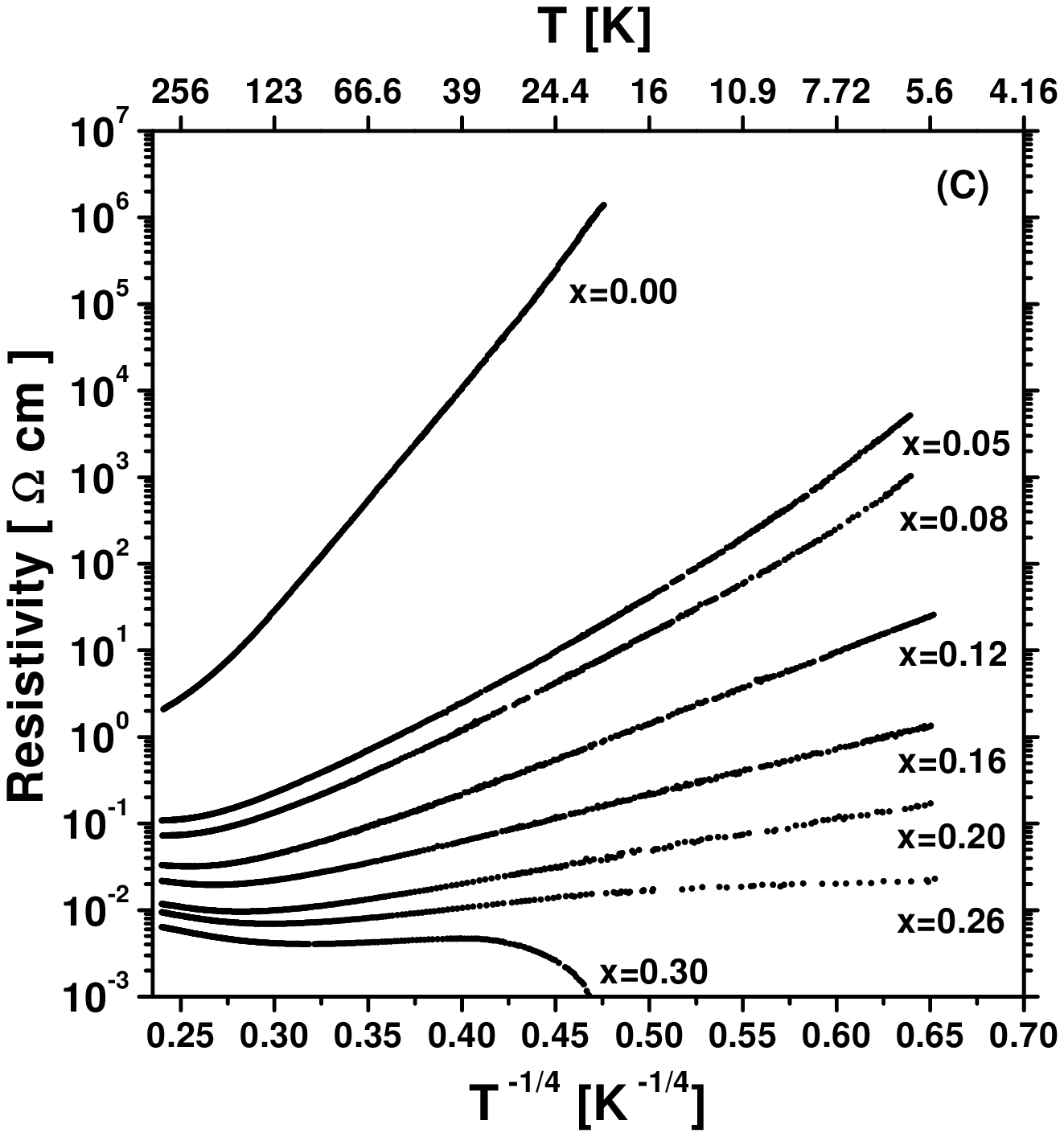,bbllx=60,bblly=380,bburx=476,bbury=800,width=6cm}}
    \vspace{0.3cm}
    \caption[fig2]{Electrical resistivity of the $Y_{1-x}Ca_{x}Ba_{2}Cu_{3}O_{6.1}$ system plotted in logarithmic scale as a function of: (A) $T^{-1/2}$, (B) $T^{-1/3}$, (C) $T^{-1/4}$.}
\end{figure}
The situation at lower temperature is more complex. For the
samples with lower $x$ ($x=0.00$, $0.05$, $0.08$) the expression
(\ref{E1}) with $n=1/2$ is preferred, indicating the VRH
conduction in the presence of carrier interactions. For the system
with $x=0.00$ this mechanism seems to occur at higher temperature
and the region below 20 K was not considered due to some anomaly
in electrical resistivity, origin of which was not clear. For the
systems with higher Ca content, lower $\chi^{2}$ values were
obtained for the relation (\ref{E1}) with $n=1/4$ signifying a VRH
conduction in 3 dimensions.  The condition (\ref{E5}) is satisfied
for the low temperature ranges of fitting for $x\leq0.2$. The low
temperature data reveal that with increasing x, a crossover from
VRH of interacting carriers to VRH of non-interacting carriers in
three dimensions occurs. This result may be universal for a
broader class of materials. Such a crossover was suggested
previously for $La_{2-y}Sr_{y}CuO_{4}$ system\cite{d45} with an
explanation that at higher $x$ the hole concentration increases
sufficiently to provide screening down to the lowest temperature.

For lower $x$ in the diagram a crossover from VRH of interacting
carriers to VRH in 3 dimensions appears with increasing
temperature. For higher $x$ temperature induces a crossover in
dimensionality. The question is why 2-dimensional hopping exists
at higher temperature and is transformed to 3-dimensional with
lowering the temperature. The $Y_{1-x}Ca_{x}Ba_{2}Cu_{3}O_{6.1}$
systems are regarded as quasi two-dimensional and the conductivity
is principally related to the Cu-O planes. In consequence, it is
not surprising that VRH in 2 dimensions is observed. At lower
temperatures it is possible that the tunneling between the planes
can compete with the motion within the planes. A crossover from
2-dimensional behavior to 3-dimensional at lower temperature have
already been observed in some quasi two-dimensional
materials\cite{Valla}.

In a case of the systems with higher level of doping the condition
(\ref{E5}) excludes the existence of VRH in some regions. For the
sample with $x=0.20$ the resistivity between 4 K and 30 K can be
interpreted as a hopping in three dimensions. However, the range
of 30 K to 70 K remains controversial. A breakdown of the VRH
conductivity is well visible for the sample with $x=0.26$. In the
considered temperature range of 20 K to 50 K the fit to the
relation (\ref{E2}) describing the conductivity in disordered
metals is characterized by the smallest $\chi^{2}$ value. As the
coefficient $\sigma_{0}$ equals zero within experimental error,
the relation (\ref{E2}) also corresponds to the predictions of the
Luttinger liquid (\ref{E4}) with $\alpha=3/8$. For this sample the
best fit among different types of hopping yielded the value of
$T_{0}'''=8.9\cdot10^{2}$ K for the VRH in 3 dimensions fitted in
the range 20 K to 50 K. This value also excludes the existence of
the VRH conductivity in the considered ranges of temperature. The
resistivity at higher temperature is not discussed as the
transition to the regime of positive $\rho(T)$ slope is quite
close. At lower temperature a presence of some superconducting
phase was detected in AC magnetic susceptibility\cite{Starowicz}
and therefore we restrict our consideration to the discussed
temperature range.

It was difficult to discuss resistivity mechanisms above the
transition to superconductivity for the sample with $x=0.3$. In
this case, a range of negative $\rho(T)$ slope is limited by the
transition to superconductivity on one side and by the positive
$\rho(T)$ slope regime at higher temperature. The analysis in this
restricted range seems unlikely to deliver any valuable
information. The fits for the VRH mechanisms at the temperature of
45 K to 70 K allowed to estimate the parameters $T_{0}'''=35$ K
for VRH in 3 dimensions and $T_{0}''=17$ K for VRH in 2 dimensions
and VRH should be rejected because of the condition (\ref{E5}).
\begin{figure}[c!]
    \vspace{0.6cm}
    \centerline{\epsfig{file=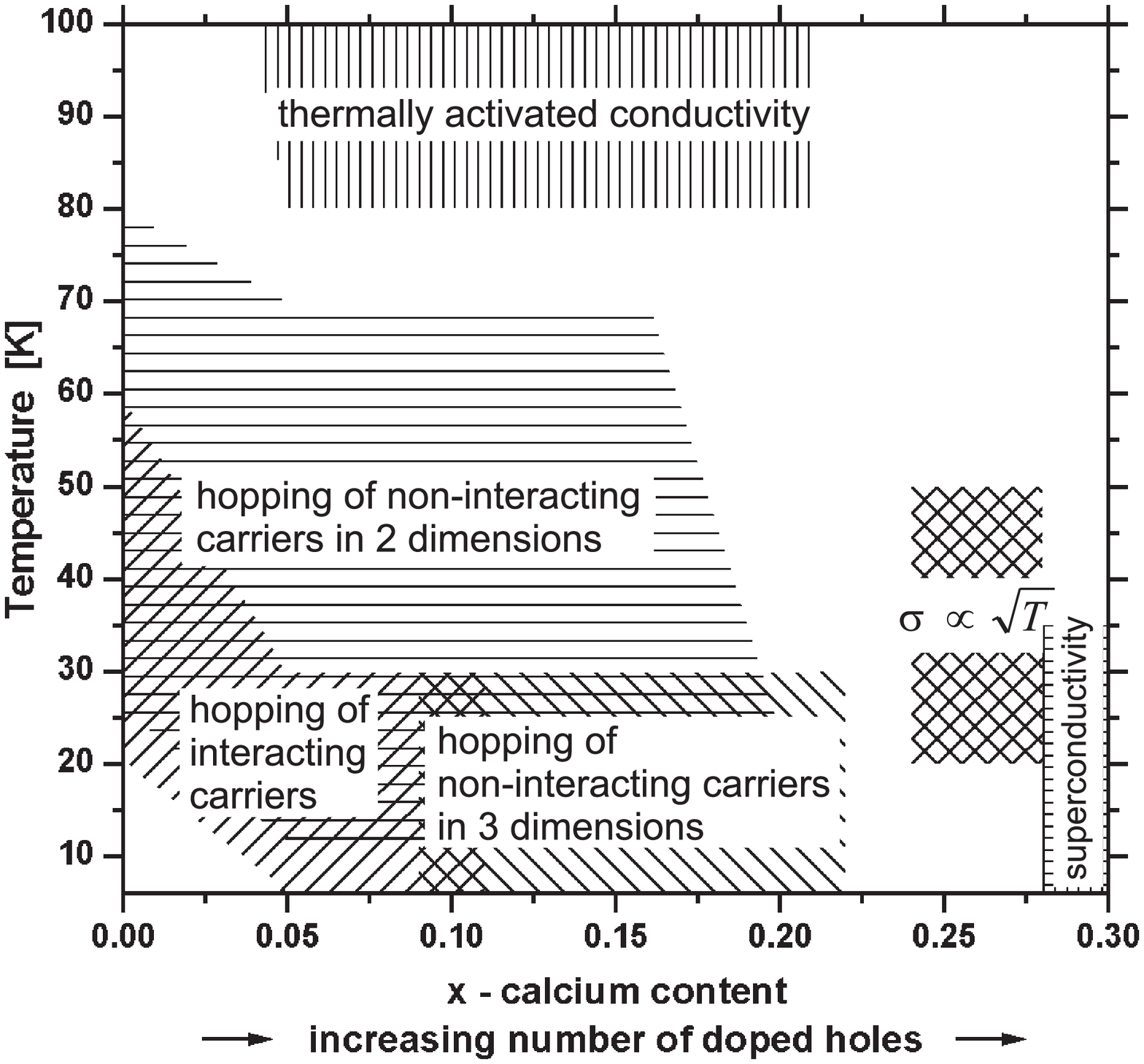,bbllx=20,bblly=250,bburx=560,bbury=760,width=8cm}}
    \vspace{0.3cm}
    \caption[fig3]{The proposed diagram of the conductivity
    mechanisms for $Y_{1-x}Ca_{x}Ba_{2}Cu_{3}O_{6.1}$ system.}
\end{figure}
At higher temperature a tendency to thermally activated
conductivity appears. Although the $\chi^{2}$ values are the
smallest for $n=1$ in these regions, the curves obtained by
$\chi^{2}$ minimisation do not fit the data perfectly. This
behavior probably cannot be described by pure relation (\ref{E1})
with $n=1$.

In summary, on the basis of systematical measurements of the
electrical resistivity we have proposed a diagram of the
conductivity mechanisms in the deoxygenated
$Y_{1-x}Ca_{x}Ba_{2}Cu_{3}O_{6.1}$ system. At the intermediate
temperature the existence of VRH in 2 dimensions is postulated, at
lower temperature a crossover from VRH of interacting carriers to
VRH of non-interacting carriers in 3 dimensions is observed with
doping. For the highly doped samples a hopping conductivity should
be excluded while the relation proposed for highly disordered
metals or for the Luttinger liquid matches the experimental data
better than the other relations do. Evidence for thermally
activated conductivity is found at higher temperature. This
diagram may be universal for the insulator-metal transitions
induced by doping in different families of the high $T_{c}$
cuprates.

The authors are grateful to T. Plackowski and Z. Tomkowicz for
help in experimental part of this work. P.S. appreciates the
discussions with Prof. J. Spa\l{}ek.
\bibliographystyle{prsty}
\bibliography{ups}


\end{multicols}

\newpage
\widetext
\begin{table}
\caption{The best fits to the experimental data for the selected
ranges of temperature. For the relation (\ref{E1}) an exponent $n$
and the values of $T_{0}'$, $T_{0}''$ and $T_{0}'''$ for VRH
conductivity are given. In a case that the condition (\ref{E5}) is
not fulfilled within the whole fitting range, the temperature
satisfying this condition is marked.}
\begin{tabular}{lccc}
Investigated sample&&Temperature range of fitting& \\
&&Best fit& \\
\tableline
$YBa_{2}Cu_{3}O_{6.16}$&20 K - 60 K&26 K - 80 K&180 K - 300 K \\
&$n=1/2$, $T_{0}'=5.6\cdot10^{3}$K&$n=1/3$,
$T_{0}''=2.4\cdot10^{5}$ K&$n=1$ \\
\tableline
$Y_{0.95}Ca_{0.05}Ba_{2}Cu_{3}O_{6.15}$&6 K - 30 K&12 K - 70 K&80 K - 200 K \\
&$n=1/2$, $T_{0}'=9.2\cdot10^{2}$ K&$n=1/3$,
$T_{0}''=2.1\cdot10^{4}$ K&$n=1$ \\
\tableline
$Y_{0.92}Ca_{0.08}Ba_{2}Cu_{3}O_{6.08}$&6 K - 30 K&12 K - 70 K&80 K - 200 K \\
&$n=1/2$, $T_{0}'=6.8\cdot10^{2}$ K&$n=1/3$,
$T_{0}''=1.6\cdot10^{4}$ K&$n=1$ \\
\tableline
$Y_{0.88}Ca_{0.12}Ba_{2}Cu_{3}O_{6.09}$&6 K - 30 K&12 K - 70 K&80 K - 200 K \\
&$n=1/4$, $T_{0}'''=1.3\cdot10^{5}$ K&$n=1/3$,
$T_{0}''=6.1\cdot10^{3}$ K&$n=1$ \\
\tableline
$Y_{0.84}Ca_{0.16}Ba_{2}Cu_{3}O_{6.09}$&6 K - 30 K&12 K - 70 K&80 K - 160 K \\
&$n=1/4$, $T_{0}'''=2.3\cdot10^{4}$ K&$n=1/3$,
$T_{0}''=1.8\cdot10^{3}$ K&$n=1$ \\
\tableline
$Y_{0.80}Ca_{0.20}Ba_{2}Cu_{3}O_{6.11}$&6 K - 30 K&12 K - 70 K&80 K - 120 K \\
&$n=1/4$, $T_{0}'''=5.3\cdot10^{3}$ K&$n=1/3$,
$T_{0}''=6.7\cdot10^{2}$ K&$n=1$ \\
&&condition (\ref{E5}): $T<25$ K& \\
\tableline
$Y_{0.74}Ca_{0.26}Ba_{2}Cu_{3}O_{6.11}$&20 K - 50 K&-&- \\
&relation (\ref{E2}): $\sigma=\sigma_{0}+m\cdot\sqrt{T}$&& \\
\end{tabular}
\end{table}
\narrowtext

\end{document}